\documentclass[a4paper,12pt]{article}
\pdfoutput=1

\usepackage{color}

\usepackage{amssymb} 
\usepackage{amsmath}
\usepackage{epsfig}
\usepackage{graphicx}

\makeatletter

\@addtoreset{equation}{section}
\makeatother
\def\be{\begin{equation}}
\def\ee{\end{equation}}
\def\bea{\begin{eqnarray}}
\def\eea{\end{eqnarray}}

\def\({\left(}
\def\){\right)}
\def\<{\left<}
\def\>{\right>}

\def\be{\begin{equation}}
\def\ee{\end{equation}}

\def\ben{\begin{eqnarray}}
\def\een{\end{eqnarray}}
\def\({\left(}
\def\){\right)}
\def\<{\left<}
\def\>{\right>}
\def\!{\right|}
\def\|{\left|}

\def\[{\left[}
\def\]{\right]}

\def\+{\bar}
\def\mb{\mathbb}

\def\A{{\cal{A}}}

\def\F{{\mathcal{F}}}

\def\O{{\cal{O}}}

\def\eps{{\cal{\varepsilon}}}

\def\F{{\cal{F}}}

\begin{document}

\setlength{\unitlength}{1mm}

\pagestyle{empty}
\vskip-10pt
\vskip-10pt
\hfill 
\begin{center}
\vskip 3truecm
{\large \bf
A reparametrization invariant 
nonabelian surface holonomy
}
\vskip 1.3truecm
{
 \textsc{Dongsu Bak, Andreas Gustavsson}}
\vspace{0.8cm} 
\begin{center} 
 Physics Department \& Natural Science Research Institute\\ 
University of Seoul, 163 Siripdaero, Tongdaemun, Seoul 02504 KOREA
\end{center}
\end{center}
\vskip0.1cm
\centerline{
\tt{(\small dsbak@uos.ac.kr, agbrev@gmail.com})
} 

\vskip 2truecm
\centerline{ABSTRACT} \vspace{0.65cm}
{\noindent 
We introduce a nonabelian surface holonomy that is constructed from a one-form gauge potential that takes values in a loop algebra of the $U(N)$ gauge group. The surface holonomy parallel transports a nonabelian string. Although it is not manifest in our formulation, we will see that our nonabelian surface holonomy is invariant under reparametrizations of the surface.}

\vfill
\vskip4pt
\eject
\pagestyle{plain}

\section{Introduction}
The question whether we can have a nonabelian generalization of the abelian Wilson surface 
\bea
W(\Sigma) &=& \exp \(i e \int_{\Sigma} B\)
\eea
for a closed surface $\Sigma$ was analyzed in \cite{Teitelboim:1985ya}. It was shown, by using the method of \cite{Teitelboim:1972vw}, that a nonabelian generalization can not be foliation invariant. If we foliate the surface by closed loops, then as we change the foliation infinitesimally, it was shown that the Wilson surface will be invariant only if the gauge group is abelian. This argument relies on the assumption that the nonabelian generalization involves a two-form gauge field of the form $B = B^a t_a$ where $t_a$ are generators of the gauge group satisfying a Lie algebra
\bea
[t_a,t_b] &=& i f_{ab}{}^c t_c
\eea
This no-go result was circumvented in \cite{Alvarez}, and then further generalized in \cite{Baez}. 
In this approach one introduces, in addition to the nonabelian two-form $B$,  a nonabelian one-form gauge potential $A$, thereby obtaining a nonabelian theory of both lines and surfaces. 
In order for this construction to work we have to impose a flatness condition, the vanishing of the so-called fake curvature, on the field strength $F$ of the one-form $A$ \cite{Baez}. While this construction is mathematically consistent, it may not be the right approach to the theory of tensionless strings (see for example \cite{Lindstrom:1990qb} for a simple example of a free tensionless string without color indices) that we expect are the degrees of freedom in the theory of multiple coincident M5 branes in M-theory. We have no Wilson lines in the theory of an abelian tensor multiplet and since lines are geometric objects, it is unnatural to expect that lines would arise only for  semisimple nonabelian gauge 
algebras if they were not already there for abelian gauge groups. Another reason comes from how the vanishing fake curvature condition should be imposed if all fields transform in the adjoint representation as they probably should because of the maximal supersymmetry most likely relating all the fields to one another in such a way that we may not tolerate different representations in the supersymmetry multiplet. In that case the fake curvature condition requires 
 $F$ to be proportional to $B$. 
Reading this relation in the other direction, we see that it constrains $B$ so severely that $D B = D F = 0$ by the Bianchi identity for $F$. 
These problems suggest that we may need to abandon this approach in order to describe the theory of multiple coincident M5 branes.

Another motivation for abandoning the construction that uses lines and surfaces together, comes from the reference \cite{Gustavsson:2026qfs} where a tensile string and its motion over a lattice discretization of Euclidean spacetime were taken as a starting point to construct a surface holonomy that is associated with the transformation of the string wave function as the string moves across the lattice. It was argued that we need an abundance of color indices distributed over the string and that each color index is closely tied to our ability to move the string across the lattice. This could hint towards a possibly deeper connection between color indices and spacetime. It also suggests  an alternative approach to
the surface holonomy in the continuum limit that does not rely on the construction using 
lines and surfaces together. Thus we now have 
three different reasons, 
all of which suggest that we might have to abandon this approach: 
\begin{itemize}
\item Wilson lines are absent in the abelian tensor multiplet theory.
\item The vanishing fake curvature condition,  
which implies $D B=0$ or an analogous constraint for other representations, 
appear to be too restrictive.
\item 
{Lattice formulations suggest that, in the continuum limit, we should place color indices all over the string.}
\end{itemize}
The lattice formulation of \cite{Gustavsson:2026qfs} shows that we have locality along the string in the sense that if we deform the string locally at one point $s_1$ and then deform the string locally at some different point $s_2$, then the associated surface holonomy we get will be the same as the surface holonomy that we get if we perform the local variation at $s_2$ first and then perform the deformation at $s_1$. In other words, local deformations along the string should commute. To obtain this locality along the string in the continuum limit, we will make use of a loop algebra. We will introduce a one-form gauge field $A_M^a(x)$ on spacetime and then we will map this into a one-form gauge field on loop space by the map
\bea
\A(C) &=& \int ds A_M^a(C(s)) t_a(s) \delta C^M(s)
\eea
where the generators $t_a(s)$ commute with one another for different points on the loop, and satisfy a loop algebra without central extension. For our definitions of the loop space and its one-form basis elements $\delta C^M(s)$, we refer to \cite{Bak:2025pcz}. The fact that we use a one-form $A_M^a(x)$ on spacetime might seem strange at first, since we would naively expect that we should have a nonabelian two-form on spacetime since we have a surface. For abelian gauge groups two-forms are naturally integrated over a surface and we define an abelian one-form on loop space by a transgression map from a two-form on spacetime as
\bea
\A(C) &=& \int ds B_{MN}(C(s)) \frac{dC^M(s)}{ds} \delta C^N(s)
\eea
For the nonabelian case, things appear to be quite different, where the loop algebra generators $t_a(s) ds$ take the role of the tangential derivatives $\frac{dC^M(s)}{ds} ds$ in the abelian case in the sense that they both transform as one-forms on the string. 

In this note, we will demonstrate that our Wilson surface, and more generally, surface holonomy, is invariant under reparametrizations of the surface.  

Since gauge symmetry is central to our construction, 
we begin in section \ref{string}  by presenting how the gauge symmetry transforms a wave function of a heavy closed string. Then in section \ref{hol} we present our definition of the surface holonomy that transports the string wave function gauge covariantly. In section \ref{fol} we show that the surface holonomy is invariant under deformations of the foliation of the surface. In section \ref{rep} we combine the foliation symmetry with reparametrization symmetry along each loop, to show that we have full reparametrization symmetry of the surface. The final section is devoted to concluding remarks.

\section{The nonabelian string}\label{string}
We consider a closed string that is embedded in spacetime as $s \mapsto C^M(s)$ for some embedding functions $C^M(s)$ in spacetime. We introduce an infinite-dimensional loop algebra extension of the $U(N)$ Lie algebra that is generated by loop algebra generators $t_a(s)$ that are parametrized by a continuous parameter $s$ around a closed loop and that satisfy 
\bea
[\, t_a(s),\, t_b(s')] &=& i \delta(s-s') f_{ab}{}^c\, t_c(s)\label{tasalg}
\eea
We next introduce a wave function $\psi^i(s)$, {with $i = 1,...,N$, that} transforms in the fundamental representation of $U(N)$. 
This wave function is defined on the closed string and it is parametrized by the parameter $s$. This wave function does not see the embedding of the string in spacetime. Instead, the wave function describes the embedding of the string into the complex multi-color space $\mb{C}^N$ and we may think of its squared norm as giving  the probability that the string at the parameter value $s$ is located inside the $i$-th M5 brane. We define the action of the generator on the wave function element as follows,
\bea
t_a(s) \psi^i(s') &=& \delta(s-s') (t_a)^i{}_j \psi^j(s)\label{tas}
\eea
where on the right-hand side appears the generators $(t_a)^i{}_j$ of the $U(N)$ Lie algebra, satisfying 
\bea
[t_a,t_b] &=& i f_{ab}{}^c t_c\label{sat}
\eea
One may check that (\ref{tasalg}) is realized on these wave function elements by using (\ref{tas}). We have an explicit realization of (\ref{tas}) as
\bea
t_a(s) &=& \(t_a\)^i{}_j \psi^j(s) \frac{\delta}{\delta \psi^i(s)}
\eea
It is interesting to note that if we attempted to introduce a wave function of the form $\psi^i(C(s))$, we would be unable to realize the loop algebra generators and their action on the wave function correctly. The loop algebra requires a factor of $\delta(s-s')$, whereas a wave function whose dependence on $s$ arises through $C(s)$ would instead give rise to $\delta^6(C(s)-C(s'))$, and hence would not yield the desired realization of the loop algebra. We interpret this observation as indicating that the spacetime embedding of the string should not be implemented by including the spacetime coordinate explicitly as an argument of the string wave function. Rather, it should be encoded indirectly through the color indices. This idea was realized explicitly in the lattice construction of \cite{Gustavsson:2026qfs}, where continuum spacetime was replaced by a lattice and a color index was assigned to every edge of the lattice. A closed string is represented by a closed sequence of edges together with the corresponding sequence of color indices. Since these indices are distributed throughout the lattice, they effectively serve as address labels. Thus, the distribution and ordering of these indices determine the location and shape of the string on the lattice, without requiring additional continuous spacetime coordinates. Extending this picture directly to continuum spacetime is difficult, however, because the continuum contains uncountably many points, making it impossible to assign a distinct discrete color label to every point in the same manner. Nevertheless, the lattice construction provides the intuitive picture underlying our approach.

We define a $U(N)$ gauge group element $g(C)$ as
\bea
g(C) &=& \exp \(i e \int ds \Lambda^a(C(s)) t_a(s)\)\label{gofC}
\eea
Here $\Lambda^a(x)$ are smooth gauge parameters defined on spacetime. No path ordering is needed in order to define this exponent unambiguously, that is, in a reparametrization invariant way, since $t_a(s)$ commute at any two different points along the loop and $t_a(s) ds$ transforms like a covariant one-form. 

The group element acts on the wave function as
\bea
g(C) \psi^i(s) &=& g^i{}_j(C(s)) \psi^j(s)
\eea
where we define 
\bea
g^i{}_j(C(s)) &=& \(e^{i e \Lambda^a(C(s)) t_a}\){}^i{}_j
\eea
Now this can be generalized to an arbitrary representation. All that we need to do, is to replace $i$ with an index $\alpha$ that transforms in the given representation. Then starting out from 
\bea
t_a(s) \psi^{\alpha}(s') &=& \delta(s-s') (t_a)^{\alpha}{}_{\beta} \psi^{\beta}(s)
\eea
we derive 
\bea
g(C) \psi^{\alpha}(s) &=& g^{\alpha}{}_{\beta}(C(s)) \psi^{\beta}(s)
\eea
For an integrated wave function over an interval $I = [s_1,s_2]$ inside the closed string, this implies that its gauge transformation is given by
\bea
g(C) \int_I ds \, \psi^{\alpha}(s) &=& \int_I ds g^{\alpha}{}_{\beta}(C(s)) \psi^{\beta}(s)
\eea
Here we have brought $g(C)$ outside the integral as it does not depend on the parameter $s$. Now we would like to generalize in another direction. We consider a wave function that is a product,
\bea
\psi^{\alpha_1 \alpha_2\cdots \alpha_K}(s_1,s_2,\dots,s_K) &=& \psi^{\alpha_1}(s_1) \psi^{\alpha_2}(s_2)\cdots \psi^{\alpha_K}(s_K)\label{pr}
\eea
Its gauge transformation is 
\bea
g(C) \psi^{\alpha_1 \alpha_2\cdots \alpha_K}(s_1,s_2,\dots,s_K) &=& g^{\alpha_1}{}_{\beta_1}(C(s_1)) \cdots g^{\alpha_K}{}_{\beta_K}(C(s_K)) \cr
&& \psi^{\beta_1 \dots \beta_K}(s_1,s_2,\dots,s_K)
\eea
We can now also consider a general wave function of a hybrid form 
\bea
\psi(C) &=& \prod_{a=1}^K \int_{I_a} ds_a\, \psi^{i_a}(s_a)\label{health}
\eea
where we divide the closed string into segments $I_a$ for $a=1,...,K$ where $K = 1,2,3,\cdots$. This provides a large set of wave functions although none of them in isolation covers the continuum limit. 
By acting on the wave functions in (\ref{health}) we can show that the group composition rule
\bea
g_1(C) g_2(C) &=& g_3(C)
\eea
gets inherited from the group composition rule of the corresponding group elements in Yang-Mills theory,
\bea
g_1(x) g_2(x) &=& g_3(x)
\eea
where $g(x) = e^{i e \Lambda^a(x) t_a}$. Two gauge parameters $\Lambda_1$ and $\Lambda_2$ are mapped into a third gauge parameter $\Lambda_3$ in some rather complicated way. From this computation of $\Lambda_3$ from $\Lambda_{1,2}$ in the Yang-Mills case, we immediately obtain the structure of the $g_3(C)$ loop group element, 
\bea
e^{i e \int ds \Lambda_1^a(C(s)) t_a(s)} e^{i e \int ds' \Lambda_2^b(C(s')) t_b(s')} &=& e^{i e \int ds \Lambda_3^a(C(s)) t_a(s)}
\eea
Namely, $\Lambda_{1,2,3}(C(s))$ are nothing but the gauge parameters in the Yang-Mills case evaluated at a point $C(s)$ on the string, so the mapping from $\Lambda_{1,2}$ to $\Lambda_3$ will also be the same as in the Yang-Mills case. 
As a by-product, we have now essentially shown that the elements $g(C)$ obey a group composition law: multiplying two such elements gives a new element of the same type.
Strictly speaking, what we have shown is that whenever $g_1(C) g_2(C)$ acts on a wave function of the product form (\ref{pr}) for any $K$ then we have the group property.
But does that mean that we always have the group property? In fact it is not very difficult to prove the group property in general using the Baker–Campbell–Hausdorff formula. To second order we find that
\bea
\Lambda_3^c(C(s)) &=& \Lambda_1^c(C(s)) + \Lambda_2^c(C(s)) - \frac{e}{2} f_{ab}{}^c \Lambda_1^a(C(s)) \Lambda_2^b(C(s)) + \O(e^2)
\eea
with a similar local structure that extends to all higher orders.

\section{The nonabelian surface holonomy}\label{hol}
A surface holonomy $U(C,C_0)$ parallel transports the string from an initial string configuration $C_0$ to a final string configuration $C$ by acting on the wave function of the string as
\bea
\psi(C_0) \rightarrow \psi(C) = U(C,C_0) \psi(C_0)
\eea
A gauge transformation acts on the wave function and on the surface holonomy together as
\bea
\psi(C) &\rightarrow & \psi^g(C) = g(C) \psi(C)\cr
U(C,C_0) &\rightarrow & U^g(C,C_0) = g(C) U(C,C_0) g(C_0)^{-1}
\eea
To construct the surface holonomy explicitly, we will assume that we are  given a gauge potential $A_M^a(x)$ on spacetime that takes values in the adjoint representation of the $U(N)$ gauge group. We define a gauge field on a closed string $C$ as 
\bea
\A(C) &=& \int ds \; A_M^a(C(s)) \delta C^M(s) t_a(s) 
\eea
This gauge field is invariant under a reparametrization of the loop $C$. We explain this in detail in the Appendix.

We shall now foliate the surface with a family of closed strings that are parametrized by a time variable $t$ 
\bea
t \mapsto C^M(t)
\eea
and we will then express a point on a given loop as $X^M(t,s) := C^M(t)(s)$. We define the time component of the gauge potential as
\bea
\A_t(C(t)) &=& \int ds \; A_M^a(X(t,s)) \frac{\partial X^M(t,s)}{\partial t} t_a(s)
\eea
We define the surface holonomy $U(C(t),C(t_0);\Sigma)$ from an initial string configuration $C(t_0)$ at an initial time $t_0$ to a string configuration $C(t)$ at time $t$ along a surface $\Sigma$ connecting the two strings, by the transport equation 
\bea
\frac{d U(C(t),C(t_0))}{dt} &=& i e \A_t(C(t)) U(C(t),C(t_0))\label{defU}
\eea
together with the initial condition $U(C(t_0),C(t_0)) = 1$. The solution is a path-ordered exponent
\bea
U(C(t),C(t_0)) &=& P \exp \(i e \int_{t_0}^t dt \int ds A_M^a(X(t,s)) t_a(s) \frac{\partial X^M(t,s)}{\partial t}\)
\eea
Notice that the tangential derivative along each loop, $\frac{\partial X^M(t,s)}{\partial s}$ at a given time $t$, does not appear in our surface holonomy, and what appears instead are the internal generators $t_a(s)$. 

From the gauge transformation 
\bea
U(C(t),C(t_0);\Sigma) \rightarrow U^g(C(t),C(t_0);\Sigma) = g(C(t)) U(C(t),C(t_0);\Sigma) g^{-1}(C(t_0))
\eea
of the surface holonomy, we derive the gauge transformation
\bea
\A(C) \rightarrow \A^g(C) = g(C) \A(C) g^{-1}(C) - \frac{i}{e} dg(C) \; g^{-1}(C)
\eea
of the gauge potential for an arbitrary closed string $C$. Here the differential operator is defined as
\bea
d &=& \int ds \delta C^M(s) \frac{\delta}{\delta C^M(s)}
\eea
If we are given a time-parametrized family of closed loops $t\mapsto C(t)$, then for the time-component of the gauge field, we have a very similar transformation rule, 
\bea
\A_t^g(C(t)) &=& g(C(t)) \A_t(C(t)) g^{-1}(C(t)) - \frac{i}{e} \frac{d g(C(t))}{d t} g^{-1}(C(t)) 
\eea
For an infinitesimal gauge variation, we may expand the gauge parameter to first order,
\bea
g(C) &=& 1 + i e \Lambda(C) + \O(\Lambda(C)^2)
\eea
We then obtain the infinitesimal gauge variation
\bea
\delta_\Lambda \A(C) = d \Lambda(C) - i e [\A(C),\Lambda(C)] + \O(\Lambda(C)^2)
\eea
We may then also define a gauge covariant derivative as
\bea
D &=& d - i e \A(C)
\eea
that acts in a certain representation of the loop algebra, as determined by the object on which it acts. For an infinitesimal gauge parameter $\Lambda$ we then have the infinitesimal gauge variation of the gauge field
\bea
\delta_\Lambda \A = D \Lambda = d\Lambda - i e [\A,\Lambda]
\eea
which explicitly becomes
\bea
\delta_\Lambda \A &=& \int ds \(\partial_M \Lambda^a + e A_{M}^b \Lambda^c f_{bc}{}^a\) \delta C^M t_a(s)
\eea
The gauge algebra can be obtained from this infinitesimal version of the gauge variation by making a second variation and then by commuting the order of these two variations. In this way we get
\bea
[\delta_{\Lambda'},\delta_{\Lambda}] \A &=& D \Lambda''
\eea
where the gauge parameter is 
\bea
\Lambda'' &=& - i e [\Lambda',\Lambda]
\eea
Its explicit form is 
\bea
\Lambda''(C) &=& e \oint ds \Lambda'^b(C(s)) \Lambda^c(C(s)) f_{bc}{}^a t_a(s)
\eea
For the finite gauge transformations, they form a group. If we make a second gauge transformation,\footnote{There are two different views that one may have on how two consecutive gauge transformations shall act. The first view (and perhaps the most natural view) is that the second gauge transformation acts on $A^g$ and transforms it into $(A^g)^{g'}$. The second view (that we adopt here) is that the second gauge variation again acts on $A$ which then leads to  $A^g$ to transform into $(A^{g'})^g$.} we find that
\bea
(\A^{g'})^{g} &=& \A^{gg'}
\eea
The gauge covariant field strength is 
\bea
\F &=& d\A - i e \A \wedge \A
\eea
which transforms as
\bea
\F^g &=& g \F g^{-1}
\eea
The explicit form of the field strength is 
\bea
\F(C) &=& \frac{1}{2} \int ds F_{MN}^a(C(s)) \delta C^M(s) \wedge \delta C^N(s) t_a(s)
\eea
where 
\bea
F_{MN}^a &:=& \partial_M A_N^a - \partial_N A_M^a + e A_M^b A_N^c f_{bc}{}^a\label{FMN}
\eea

\section{Foliation independence of the surface holonomy}\label{fol}
From the defining equation (\ref{defU}), it follows that
\bea
\frac{d}{dt} \(U^{-1}(t,t_0) \delta U(t,t_0)\) &=& i e U^{-1}(t,t_0) \delta \A_t(t) U(t,t_0)
\eea
where, in this section, we are using a shorthand notation $U(t,t_0)$, for the surface holonomy $U(C(t),C(t_0);\Sigma)$. Let us now integrate the above relation from the initial time $t_0$ up to some final time $t_1$,
\bea
U^{-1}(t_1,t_0) \delta U(t_1,t_0) &=& i e \int_{t_0}^{t_1} dt U^{-1}(t,t_0) \delta \A_t(t) U(t,t_0)
\eea
where we have noted that $U(t_0,t_0) = 1$, whose variation vanishes. 

For the right-hand side we need to obtain how $\A_t$ varies as we make an infinitesimal variation of the loop $C$ at a given time $t$, 
\bea
\delta C^M(t,s) &=& \xi^M(t,s)
\eea
The variation is computed as $\delta \A_t(C) := \A_t(C+\xi) - \A_t(C)$,
so it transforms as a scalar under this specific variation of the loop, in which 
we do not change the parametrization of the time variable $t$. We compute this variation with the result
\bea
\delta \A_t(C) &=& \int ds \(\partial_M A_N^a - \partial_N A_M^a\) \xi^M \frac{\partial C^N}{\partial t} t_a(s) + \frac{d}{dt} \int ds \(A_M^a \xi^M\) t_a(s)
\eea
To make the notation more convenient, we will now absorb $t_a(s)$ into the gauge field and use a shorthand notation as $A_M := A_M^a(C(t,s)) t_a(s)$ and similarly, for the field strength we will write it as $F_{MN} := F_{MN}^a(C(t,s)) t_a(s)$ where we define $F_{MN}^a$ as in equation (\ref{FMN}). We then get 
\bea
U^{-1}(t_1,t_0) \delta U(t_1,t_0) &=& i e \int_{t_0}^{t_1} dt U(t,t_0)^{-1}\int ds \(\partial_M A_N - \partial_N A_M\) \xi^M \frac{\partial C^N}{\partial t}   U(t,t_0)\cr
&& + i e \left[\int ds U^{-1} A_M U \xi^M \right]^{t_1}_{t_0}\cr
&& + i e \int_{t_0}^{t_1} dt U^{-1}\frac{dU}{dt} \int ds U^{-1} A_M U  \xi^M \cr
&& - i e \int_{t_0}^{t_1} dt \int ds U^{-1} A_M \frac{dU}{dt}\xi^M
\eea
The last two lines, when we apply the defining equation (\ref{defU}), give us a commutator, 
\bea
&& - e^2 \int_{t_0}^{t_1} dt U^{-1} \left[\A_t,\int ds A_M \xi^M\right] U\cr
 &=& i e \int_{t_0}^{t_1} dt U^{-1}\int ds e A_{M}^a A_{N}^b f_{ab}{}^c t_c(s) \xi^M \frac{\partial C^N}{\partial t} U
\eea
This then combines with the remaining linear terms into the field strength tensor, giving 
\bea
U^{-1}(t_1,t_0) \delta U(t_1,t_0) &=& i e \int_{t_0}^{t_1} dt U^{-1}\int ds F_{MN} \xi^M \frac{\partial C^N}{\partial t}   U\cr
&& + i e \left[\int ds U^{-1} A_M \xi^M U \right]^{t_1}_{t_0}
\eea
We now multiply both sides by $U(t_1,t_0)$ and apply the relation
\bea
U(t_1,t_0) U^{-1}(t,t_0) &=& U(t_1,t)
\eea
The final result is
\bea
\delta U(t_1,t_0) &=& i e \int_{t_0}^{t_1} dt U(t_1,t) \int ds F_{MN}(t) \xi^M(t)\frac{\partial C^N(t)}{\partial t} U(t,t_0)\cr
&& + i e \left[\int ds U(t_1,t) A_{M}(t)\xi^M(t) U(t,t_0)  \right]^{t_1}_{t_0}\label{bndry}
\eea
We now let the variation of the loop at each time be given by
\bea
\xi^M(t,s) &=& \eps(t,s) \frac{\partial X^M(t,s)}{\partial t}\label{xi}
\eea
where we impose Dirichlet boundary conditions at the initial and final time,
\bea
\eps(t_1,s) &=& 0\cr
\eps(t_0,s) &=& 0
\eea
which means that the boundary terms vanish, and we get only a contribution from the points $t$ in the open interval $t_1 > t > t_0$. But there we see that 
the contraction of the antisymmetric tensor $F_{MN}$ with the symmetric combination $\partial_t C^M \partial_t C^N$ vanishes. 
So we conclude that, with this particular variation, we get a vanishing variation of the surface holonomy, 
\bea
\delta U(t_1,t_0) &=& 0
\eea
This means that the surface holonomy is invariant under transverse deformations of the loops that are foliating the surface, while we keep the initial and final loops fixed. That is to say that we keep the geometric shape of the surface fixed while we change how it is foliated by loops.

\section{Reparametrization of the surface}\label{rep}
Our surface is parametrized by two parameters, $s$ and $t$. Let us first study the deformation
\bea
\delta X^M(t,s) &=& \eps(t,s) \frac{\partial X^M(t,s)}{\partial t}
\eea
We view the $X^M$ as $D$ scalar fields on the surface. As such, they transform as 
\bea
X'^M(t',s') &=& X^M(t,s)
\eea
Let us make an infinitesimal reparametrization of the surface of the form
\bea
t' &=& t - \eps(t,s)\cr
s' &=& s 
\eea
Then 
\bea
\delta X^M(t,s) &=& \eps(t,s) \frac{\partial X^M(t,s)}{\partial t}
\eea
This is exactly the variation that we considered above, for which we showed that the surface holonomy is invariant, if we also impose Dirichlet boundary conditions at the initial and final time. After the deformation, the constant time slices are
\bea
(t',s') &=& (t - \eps(t,s),s)
\eea
for constant $t$. 

But we also have reparametrization invariance along each constant time slice. So we can change the parameter $s$ to another parameter $s_1$ on each constant time slice. That means that we transform the initial coordinates as
\bea
(t,s) \rightarrow (t,s_1)
\eea
where we define, for each constant $t$ and arbitrary $s$ along that constant $t$ slice, a new variable $s_1$ along that slice such that
\bea
s &=& f(t,s_1)\cr
s_1 &=& h(t,s)
\eea
where the latter is the inverse relation to the former and both $f$ and $h$ are smooth nondegenerate functions of two variables. Still, also for the new parameter $s_1$, the constant time slices will be for constant $t$.

Now let us again make the transformation 
\bea
(t',s') &=& (t - \eps(t,f(t,s_1)),f(t,s_1))
\eea
Here $\eps(t,s)$ is arbitrary so if we put 
\bea
g(t,s_1) &=& t - \eps(t,f(t,s_1))
\eea
then $g(t,s)$ is also arbitrary, subject to the Dirichlet boundary conditions
\bea
g(t_0,s_1) &=& t_0\cr
g(t_1,s_1) &=& t_1
\eea
We have obtained a general reparametrization of the form
\bea
t' &=& g(t,s_1)\cr
s' &=& f(t,s_1)
\eea
relating the old parameters $(t,s_1)$ to the new parameters $(t',s')$. Since both $f$ and $g$ are arbitrary functions of the old parameters, subject to suitable boundary conditions and a nonvanishing Jacobian, we see that we have now obtained a rather general reparametrization of the surface. 

It remains to understand whether we can extend the reparametrization symmetry to allow for a general reparametrization of the surface at a boundary, say the final boundary at $t=t_1$. To this end, let us start by noting that if we do not impose the Dirichlet condition at this boundary, then we need to keep the boundary term in the variation of the surface holonomy, which from (\ref{bndry}) and (\ref{xi}), becomes
\bea
\delta U(t_1,t_0) &=& i e \int ds A_{M}(t_1,s) \xi^M(t_1,s) U(t_1,t_0) \label{var0}
\eea
This is nothing but a gauge variation of the surface holonomy with the gauge parameter $\Lambda = A_M \xi^M$ where $\xi^M(t,s) = \eps(t,s) \frac{\partial X^M(t,s)}{\partial t}$. But now as we lifted the Dirichlet boundary condition, the surface itself has been slightly deformed at the boundary, by the amount
\bea
\delta X^M(t_1,s) &=& \eps(t_1,s) \frac{\partial X^M(t_1,s)}{\partial t}
\eea 
While this allowed us to perform a general reparametrization locally around $t=t_1$ it thus came with the cost of creating a physical deformation of the geometric surface, which is not just a reparametrization of the surface. It is a physical deformation at the boundary. Now to conclude that we can have a general reparametrization of the surface at the boundary, we need to demonstrate that the surface holonomy is exactly invariant, 
rather than merely invariant up to a gauge transformation, 
if we do not physically deform the geometry of the surface. In order to answer that question, we need to understand exactly how the physical deformation of the surface changes the surface holonomy and then we can subtract that variation to obtain the surface holonomy of the original geometric surface, now with a general reparametrization all the way out to its boundary at $t=t_1$. This step is surprisingly easy. When we physically deform the surface by the amount $\delta X^M(t_1,s) = \xi^M(t_1,s)$, then our surface holonomy changes by the amount
\bea
\delta_{geom} U(t_1,t_0) &=& i e \int ds A_{M}(t_1,s) \xi^M(t_1,s) U(t_1,t_0) 
\eea
This result is independent of how we parametrize the surface as it only depends on the variation $\delta X^M = \xi^M$ of the embedding coordinates of the surface in spacetime. Now if we subtract this geometric variation from the reparametrization variation in (\ref{var0}), the net result is a vanishing variation of the surface holonomy. By a very similar argument one can also understand that we have general reparametrization invariance at the initial boundary at $t = t_0$. We conclude that the surface holonomy is fully reparametrization invariant.

\section{Discussion} 
We have shown that we get a fully reparametrization-invariant surface holonomy when we make the generalization from an abelian to a nonabelian gauge group by replacing the integrand $B_{MN}\frac{dC^M}{ds}\,ds $ with $ A_N^a t_a(s)\,ds$. We motivated this replacement by observing that both integrands transform as one-forms under a reparametrization of the loop parameter $s$. Since $t_a(s)$ and $t_b(s')$ commute for different values of $s$ and $s'$, the surface holonomy acts locally along the string. This statement, however, should not be interpreted as requiring the underlying string to be tensionless. The surface holonomy could instead arise from a tensile string inside a stack of coincident M5-branes. Even if the tension is very large, or formally infinite, moving only one portion of the string produces, according to our definition, only a local contribution to the surface holonomy.

This is precisely what one should expect. The interaction transmitted along the string by its tension belongs to the dynamical part of an effective action, for example to a Nambu–Goto-type term, whereas the surface holonomy is a topological factor. The tension therefore need not enter the surface holonomy itself. Just as a Wilson line is independent of the mass of the quark to which it is attached, the same surface holonomy can consistently accompany strings with different tensions, from a tensionless string to an arbitrarily heavy one. Changing the tension changes the dynamics of the string, but not its topological parallel transport. Thus, the commutativity at different values of $s$ expresses the locality of the holonomy density and does not imply the absence of tensile interactions in the complete string theory.

\subsection*{Acknowledgement}
D.B. was supported by the 2026 Research Fund of the University of Seoul.

\appendix

\section{Reparametrization of the string}
In this appendix we will show that for any Lie algebra valued spacetime field $f^a(x)$, the corresponding loop algebra valued integral is reparametrization-invariant in the sense that
\bea
\int ds f^a(C(s)) t_a(s) = \int ds' f^a(C'(s')) t'_a(s')
\eea
where $s' = s'(s)$ is a new parameter satisfying 
\bea
\frac{ds'}{ds} > 0\label{mono}
\eea
at each point $s$ on the loop. This reparametrization symmetry is true regardless of how $f^a(x)$ transforms under spacetime diffeomorphisms. We could, for example, have 
\bea
f^a(C(s)(t)) = A_M^a(C(s)(t)) \frac{\partial C^M(s)(t)}{\partial t}
\eea
where $X^M(t,s) = C^M(s)(t)$. The reparametrization invariance follows immediately if we assume that
\bea
t'_a(s') ds' &=& t_a(s) ds\label{tasapp}
\eea
together with coordinate functions transforming as scalar fields under the reparametrization, 
\bea
C'^M(s') &=& C^M(s)\label{sc}
\eea
Let us also comment that the transformation of $t_a(s) ds$ as a one-form together with the assumption (\ref{mono}) implies that the loop algebra is invariant. To see that, we use 
\bea
\delta(s'_1-s'_2) &=& \frac{\delta(s_1 - s_2)}{\left|\frac{ds'}{ds}\right|} 
\eea

Let us also obtain the infinitesimal version of the reparametrization symmetry where we take $s' = s + \eps(s)$. Since $s$ is an integration variable that we can freely rename, we have\footnote{We also need to adjust the range over which we integrate accordingly, thus if the range is $[s_0,s_1]$ for $s$, then the corresponding range will be $[s'_0,s'_1]$ for $s'$. Of course here $s_1$ and $s_0$ are periodically identified through periodic boundary conditions, and these boundary conditions need to also be adjusted accordingly as we change the parameter so that for example $t_a(s_1) = t_a(s_0)$ is transformed into $t'_a(s'_1) = t'_a(s'_0)$ and $C(s_1) = C(s_0)$ is transformed into $C'(s'_1) = C'(s'_0)$.}
\bea
\int ds' f^a(C'(s')) t'_a(s') = \int ds f^a(C'(s)) t'_a(s)
\eea
By Taylor expanding and using (\ref{tasapp}) and (\ref{sc}) we get
\bea
C'^M(s) &=& C^M(s) - \eps(s) \frac{dC^M(s)}{ds} + \O(\eps^2)\cr
f^a(C'(s)) &=& f^a(C(s)) - \eps(s) \frac{df^a(C(s))}{ds} + \O(\eps^2)\cr
t'_a(s) &=& t_a(s) - \frac{d}{ds}\(\eps(s) t_a(s)\)  + \O(\eps^2)
\eea  
By pointwise subtraction, we then get a total derivative
\bea
f^a(C'(s)) t'_a(s) - f^a(C(s)) t_a(s) &=& - \frac{d}{ds} \Big(\eps(s) f^a(C(s)) t_a(s)\Big)
\eea
and therefore, for a closed string,
\bea
\int ds \Big(f^a(C'(s)) t'_a(s) - f^a(C(s)) t_a(s)\Big) &=& 0
\eea
which is the infinitesimal statement of the reparametrization invariance.

\end{document}